\documentstyle[12pt,a4]{article}

\begin{document}

\title{Entropy flux and Lagrange multipliers: information theory and
thermodynamics  
\thanks{%
This paper is dedicated to Prof. J. Casas-V\'{a}zquez in his 60th birthday}}
\author{Raquel Dom\'\i nguez-Cascante \thanks{%
Electronic address: raquel@ulises.uab.es} and David Jou \thanks{%
Also at Institut d'Estudis Catalans, C/ Carme, 47, Barcelona} \\
Departament de F\'{\i}sica\\
Universitat Aut\`onoma de Barcelona\\
08193 Bellaterra (Barcelona), Spain}
\date{}
\maketitle

\begin{abstract}
We analyze the relation between the nonequilibrium Lagrange multipliers used
in information theory and those used in Liu's technique to exploit the
entropy inequality. In particular, we deal with some of the subtleties found
in the analysis of the entropy flux.
\end{abstract}

\section{Introduction}

Lagrange multipliers play a central role in information theory and
statistical mechanics, not only as a mathematical tool to introduce a set of
restrictions on the distribution function, but also because of their general
physical meaning as intensive parameters. When one tries to use this
technique in nonequilibrium situations, the mathematical extension is not so
complicated, in contrast with the subtle conceptual problems related to the
physical interpretation of the nonequilibrium Lagrange multipliers.

Furthermore, Lagrange multipliers are also used in nonequilibrium
thermodynamics in the elegant method proposed by Liu\cite{Liu} to take into
account the restrictions placed on the thermodynamic fields by the balance
equations. Some years ago, Dreyer\cite{???} showed that, under given
conditions, the Lagrange multipliers used in statistical mechanics and those
appearing in the formulation of nonequilibrium thermodynamics could be
identified. This is of course an important question in order to relate
microscopic and macroscopic results.

Here, we deal with some subtleties which arise in the analysis of the
entropy flux. Some of these subtleties, related to the role of
(nonequilibrium) absolute temperature \cite{Temp1,Temp2} in the definition
of the entropy flux, were considered in \cite{Cuca}, both from the
macroscopic point of view of Extended Irreversible Thermodynamics (EIT) \cite
{Ext}-\cite{Nettleton} and considering an information theoretical formalism.
In the first case, an entropy flux of the form 
\begin{equation}
\vec{J}^S=\frac{\vec{q}}\theta  \label{flux-EIT}
\end{equation}
was assumed, where $\vec{q}$ is the heat flux and $\theta $ the
nonequilibrium temperature defined as the derivative of the generalized,
flux-dependent specific entropy $s(u,\vec{q})$ considered by EIT, with
respect to the internal energy $u$. When this expression for the entropy
flux is considered, one is led to an entropy production $\sigma ^S$ of the
form 
\begin{equation}
\sigma ^S=\vec{q}\cdot \nabla \left( \frac 1\theta \right) -\frac{\vec{\alpha%
}}\theta \cdot \stackrel{.}{\vec{q}}\geq 0,  \label{ineq}
\end{equation}
where $\vec{\alpha}:=-\left( \theta /v\right) \left( \partial s/\partial 
\vec{q}\right) $and $v=1/\rho $ is the specific volume of the system whose
density is $\rho $. The simplest evolution equation for the heat flux
verifying the inequality (\ref{ineq}), would be a generalized
Maxwell-Cattaneo equation, namely 
\begin{equation}
\mu \vec{q}=\nabla \left( \frac 1\theta \right) -\frac \alpha \theta 
\stackrel{.}{\vec{q},}\qquad \mu >0,  \label{Max-Cat}
\end{equation}
with $\vec{\alpha}=:\alpha \vec{q}$ (this should hold according to the
representation theorem, as $\vec{q}$ is the only vector among the
independent variables and we should require that the equations reflect the
isotropy of the system).

However, one could add new terms to equation (\ref{flux-EIT}), what would
merely modify the expression for the entropy production and, within this
formalism, the allowed evolution equations for the heat flux verifying the
second law of Thermodynamics. For instance, by using an information
theoretical formalism a generalized entropy flux has been obtained in \cite
{Cuca}, namely, 
\begin{equation}
\vec{J}^S=\frac{\vec{q}}\theta -\frac{\vec{\alpha}}\theta \cdot {\bf Q,}
\end{equation}
where ${\bf Q}$ is the flux of heat flux and $\vec{\alpha}$ and $\theta $
are related to the Lagrange multipliers introduced in the formalism. When
this entropy flux is considered, the entropy production arising from the
evolution equation for the specific entropy would be 
\begin{equation}
\sigma ^S=\rho \dot{s}+\nabla \vec{J}^S=\vec{q}\cdot \nabla \left( \frac
1\theta \right) -\frac{\vec{\alpha}}\theta \cdot \stackrel{.}{\vec{q}}%
-\nabla \left( \frac{\vec{\alpha}\cdot {\bf Q}}\theta \right) \geq 0.
\end{equation}
On the other hand, one can consider an evolution equation for the heat flux
of the form 
\begin{equation}
\stackrel{.}{\vec{q}}+\nabla \cdot {\bf Q=}\vec{\sigma}^q.  \label{evo-q}
\end{equation}
In this case, the entropy production would be 
\begin{equation}
\sigma ^S=\vec{q}\cdot \nabla \left( \frac 1\theta \right) -\frac{\vec{\alpha%
}}\theta \cdot \vec{\sigma}^q-\nabla \left( \frac{\vec{\alpha}}\theta
\right) \cdot {\bf Q}\geq 0.
\end{equation}

Let us note that (\ref{Max-Cat}) can be written in the form (\ref{evo-q}),
if it is assumed that ${\bf Q=C,}$ being ${\bf C}$ a constant tensor and 
\begin{equation}
\vec{\sigma}^q=\frac \theta \alpha \nabla \left( \frac 1\theta \right) -%
\frac{\mu \theta }\alpha \vec{q}.
\end{equation}
(what will prove not to be convenient as it implies the assumption that $%
\nabla \theta =\nabla \theta (u,\vec{q})$ instead of $\theta =\theta (u,\vec{%
q})$) or, otherwise, that 
\begin{eqnarray}
{\bf Q} &=&\frac 1\alpha {\bf U}, \\
\vec{\sigma}^q &=&-\frac{\mu \theta }\alpha \vec{q}.
\end{eqnarray}
(but now we should assume that the product $\alpha \theta $ is a constant
independent of $u$ and $\vec{q}$). However, in a general situation, it is
not possible to write (\ref{Max-Cat}) in the form (\ref{evo-q}).

Now, we try to generalize the previous results by applying the technique
developed by Liu \cite{Liu} to exploit the entropy inequality. Let us note
that, from this new point of view, we start by knowing the proper evolution
equation for the heat flux (which can be derived, for instance, from
kinetic a
theory) and try to find restrictions to the form the specific entropy and
the entropy flux may take. We will, thus, observe under what assumptions the
previous results may be recovered. In addition, we should note that the
information theory results obtained in \cite{Cuca} and Liu's technique are
seen not to be equivalent in this situation, in spite of the results in \cite
{???}. The reason will become evident in the following section.

 The plan of
the paper is as follows: In Sec. 2, we briefly introduce the fundamentals of
Liu's technique, while in Sec. 3 we apply such a technique to derive the
form of the entropy flux and entropy production when a Maxwell-Cattaneo-like
evolution equation for the heat flux is assumed. We will observe under what
further assumptions the macroscopic results in \cite{Cuca} are recovered. In
Sec. 4, however, we will consider an evolution equation of the form (\ref
{evo-q}) instead, in order to compare it with the results obtained by using
information theory. Sec. 5 is devoted to the conclusions of this paper.

\section{Liu's technique to exploit the entropy inequality}

The aim of thermodynamics is the determination of the fields $F_i$ ($i=1..N$%
) that constitute the state space ${\cal Z}$ of the system and are
governed by balance equations of the form 
\begin{equation}
\rho \frac{\partial F_i}{\partial t}+\nabla \cdot \vec{J}^{F_i}=\sigma
^{F_i},  \label{ev-equ}
\end{equation}
where $\vec{J}^{F_i}$ and $\sigma ^{F_i}$ are, respectively, the flux and
production of $F_i$ and are functions of the state space ${\cal Z}$. Such
balance equations should be supplemented with constitutive equations
relating the fluxes and the productions to the variables spanning the state
space. If the constitutive equations were known, one would be able to
explicitly solve the balance equations and obtain the {\it thermodynamic
processes. }

However, such constitutive equations are not known, so we need a method that
may help us to obtain them or, at least, to limit their form. Liu's
technique accomplishes this by the exploitation of the entropy principle in
order to obtain expressions for the constitutive equations (or, at least, to
reduce their generality). In fact, the constitutive equations must be of
such form that they verify the following principles:

\begin{description}
\begin{enumerate}
\item  the entropy principle

\item  the requirements of convexity and causality, which demand that the
field equations be symmetric hyperbolic, and

\item  the principle of relativity.
\end{enumerate}
\end{description}

The entropy principle is a local expression for the second law of
thermodynamics, written as an evolution equation for the specific entropy $%
s\left( {\cal Z}\right) $ 
\begin{equation}
\rho \frac{\partial s}{\partial t}+\nabla \cdot \vec{J}^S=\sigma ^S\geq 0.
\label{ent-ineq}
\end{equation}
The entropy production $\sigma ^S$ must be positive for any solution of the
balance equations. These requirements may be taken into account \cite{Liu}
by introducing a set of Lagrange multipliers $\lambda _i\left( {\cal Z}%
\right) $, such that 
\begin{equation}
\rho \frac{\partial s}{\partial t}+\nabla \cdot \vec{J}^S-\sum_{i=1}^N%
\lambda _i\left( \rho \frac{\partial F_i}{\partial t}+\nabla \cdot \vec{J}%
^{F_i}-\sigma ^{F_i}\right) \geq 0,  \label{ent-Lag}
\end{equation}
for {\it all continuous differentiable fields} $F_i$. By making use of the
chain rule, equation (\ref{ent-Lag}) may be seen to adopt the form 
\begin{equation}
\sum_{i=1}^N\rho \left( \frac{\partial s}{\partial F_i}-\lambda _i\right) 
\frac{\partial F_i}{\partial t}+\sum_{i=1}^N\left( \frac{\partial \vec{J}^S}{%
\partial F_i}-\lambda _i\frac{\partial \vec{J}^{F_i}}{\partial F_i}\right)
\cdot \nabla F_i+\sum_{i=1}^N\lambda _i\sigma ^{F_i}\geq 0,
\end{equation}
and, as this expression must hold for {\it any} field, the terms in brackets
must vanish so that 
\begin{equation}
\frac{\partial s}{\partial F_i}=\lambda _i,\qquad \frac{\partial \vec{J}^S}{%
\partial F_i}=\lambda _i\frac{\partial \vec{J}^{F_i}}{\partial F_i},\qquad
\forall i.
\end{equation}
Thus we can write 
\begin{equation}
ds=\sum_{i=1}^N\lambda _idF_{i,}\qquad d\vec{J}^S=\sum_{i=1}^N\lambda _id%
\vec{J}^{F_i}  \label{Gibbs-gen}
\end{equation}
and the residual inequality reads 
\begin{equation}
\sum_{i=1}^N\lambda _i\sigma ^{F_i}\geq 0.  \label{res-ineq}
\end{equation}
We should compare (\ref{Gibbs-gen})$_2$ with 
\begin{equation}
\vec{J}^S=\sum_{i=1}^N\lambda _i\vec{J}^{F_i}  \label{Inf-flux}
\end{equation}
in \cite{Cuca}. Both expressions differ, while according to \cite{???},
information theory is equivalent to Liu's technique. The reason for such
differences relies in the definition given to entropy. While in \cite{Cuca},
entropy is defined according to Boltzmann's expression, namely 
\begin{equation}
S=-\frac{k_B}{h^{3N}N!}\int f\ln f\ d\Gamma _N,  \label{ent-B}
\end{equation}
in \cite{???} it is considered that 
\begin{equation}
S=-k_B\int \left[ f\ln \left( \frac fy\right) +\frac ya\left( 1-\frac
ayf\right) \ln \left( 1-\frac ayf\right) \right] d\Gamma _N,
\end{equation}
where $a=1,-1$ for fermions and bosons respectively and $y$ is the
degeneracy of the state. Let us note that from this expression we can easily
recover (\ref{ent-B}) if $a=0$ is chosen. The entropy flux is thus given in 
\cite{???} by 
\begin{equation}
\vec{J}^S=-k_B\int \left[ f\ln \left( \frac fy\right) +\frac ya\left(
1-\frac ayf\right) \ln \left( 1-\frac ayf\right) \right] \sum_i\vec{C}%
_id\Gamma _{N-1}d\vec{c}_N,
\end{equation}
where $\vec{C}_i=\vec{c}_i-\vec{v}$ is the peculiar velocity of particle $i,$
and if we take into account that $f$ is a generalized canonical distribution
function, namely 
\begin{equation}
f=\frac y{\exp (\sum_i\lambda _iA_i)+a},
\end{equation}
we can obtain 
\begin{equation}
S=\sum_i\lambda _i<A_i>+k_B\frac ya\int \ln \left( 1+a\exp \left(
-\sum_i\lambda _iA_i\right) \right) d\Gamma _N,
\end{equation}
and 
\begin{equation}
\vec{J}^S=\sum_i\lambda _i\vec{J}^{A_i}+k_B\frac ya\int \ln \left( 1+a\exp
\left( -\sum_i\lambda _iA_i\right) \right) \sum_i\vec{C}_id\Gamma _{N-1}d%
\vec{c}_N.  \label{ent-f}
\end{equation}

Note that if we take $a=0$ in (\ref{ent-f}), we recover equation (\ref
{Inf-flux}). However, if we evaluate the differential form 
\begin{eqnarray}
d\vec{J}^S=\sum_i\lambda _id\vec{J}^{A_i}+\sum_id\lambda _i\vec{J}^{A_i}\cr %
+k_B\frac ya\sum_id\lambda _i\frac \partial {\partial \lambda _i}\int \ln
\left( 1+a\exp \left( -\sum_j\lambda _jA_j\right) \right) \sum_k\vec{C}%
_kd\Gamma _{N-1}d\vec{c}_N,
\end{eqnarray}
the last term is easily seen to be equal to $-\sum_id\lambda _i\vec{J}^{A_i}$%
, so that 
\begin{equation}
d\vec{J}^S=\sum_i\lambda _id\vec{J}^{A_i},
\end{equation}
independently of the value of $a$, as in \cite{???}. Thus, we observe that,
although in the case of bosons and fermions the results obtained from
information theory are coincident with those obtained by the exploitation of
the entropy principle, one must be careful if classical particles are
considered. This question should be further investigated as it is not
convenient, neither can be guaranteed that 
\begin{equation}
\sum_id\lambda _i\vec{J}^{A_i}=0  \label{malfet}
\end{equation}
for classical particles.

\section{Heat flux determined by a Maxwell-Cattaneo evolution equation}

Let us consider a system at rest submitted to a heat flux and verifying a
Maxwell-Cattaneo evolution equation. Thus, the two evolution equations for
the variables $u$ and $\vec{q}$ spanning the space state are 
\begin{equation}
\rho \frac{\partial u}{\partial t}+\nabla \cdot \vec{q}=0,\qquad \frac{%
\partial \vec{q}}{\partial t}+\frac 1\tau \left( \vec{q}+\lambda \nabla
\Theta \right) =0  \label{evo-equ2}
\end{equation}
where $\tau $ is the relaxation time of heat pulses, $\lambda $ the
thermal conductivity of the material and $\Theta $ the temperature of the
system (we do not specify yet whether this temperature is the nonequilibrium
temperature, as assumed in \cite{Cuca} or the local-equilibrium one, as
usually done in linear EIT: we will consider both assumptions in the
following). (\ref{evo-equ2})$_2$ coincides with (\ref{Max-Cat}) as long as
we identify 
\begin{equation}
\theta =\Theta ,\qquad \tau =\frac \alpha {\mu \Theta },\qquad \lambda
=\frac 1{\mu \Theta ^2}.
\end{equation}
The last two equalities are mere definitions of parameters as long as the
first identification is performed. Thus, (\ref{evo-equ2})$_2$ is more
general than (\ref{Max-Cat}), because we have not already assumed what
temperature $\Theta $ is. In addition, let us remark that, as commented
above, (\ref{evo-equ2})$_2$ cannot be written in the form (\ref{ev-equ}),
unless we assume that $\nabla \Theta $ is a function of the state variables.
If such assumption is taken, we can obtain 
\begin{eqnarray}
&&\rho \left( \frac{\partial s}{\partial u}-\lambda _u\right) \frac{\partial
u}{\partial t}+\rho \left( \frac{\partial s}{\partial \vec{q}}-\vec{\lambda}%
_q\right) \cdot \frac{\partial \vec{q}}{\partial t}+\frac{\partial \vec{J}^S%
}{\partial u}\cdot \nabla u \\
&&+\left( \frac{\partial \vec{J}^S}{\partial \vec{q}}-\lambda _u{\bf U}%
\right) :\nabla \vec{q}-\vec{\lambda}_q\frac \rho \tau \left( \vec{q}%
+\lambda \nabla \Theta \right) \geq 0,  \nonumber
\end{eqnarray}
and thus 
\begin{equation}
\frac{\partial s}{\partial u}=\lambda _u,\qquad \frac{\partial s}{\partial 
\vec{q}}=\vec{\lambda}_q,
\end{equation}
\begin{equation}
\frac{\partial \vec{J}^S}{\partial u}=0,\qquad \frac{\partial \vec{J}^S}{%
\partial \vec{q}}=\lambda _u{\bf U,}
\end{equation}
\begin{equation}
-\vec{\lambda}_q\cdot \frac \rho \tau \left( \vec{q}+\lambda \nabla \Theta
\right) \geq 0.
\end{equation}
In agreement with the representation theorem, as $\vec{q}$ is the only
vectorial variable in the state space, we can write: 
\begin{eqnarray}
\vec{J}^S(u,\vec{q}) &=&\varphi (u,q^2)\vec{q},\qquad \vec{\lambda}%
_q=\Lambda (u,q^2)\vec{q}, \\
s &=&s(u,q^2),\qquad \lambda _u=\lambda _u(u,q^2),
\end{eqnarray}
so the previous equations may be rewritten as 
\begin{equation}
\frac{\partial s}{\partial u}=\lambda _u,\qquad \frac{\partial s}{\partial
q^2}=\frac \Lambda 2,
\end{equation}
\begin{equation}
\frac{\partial \varphi }{\partial u}=0,\qquad \frac{\partial \varphi }{%
\partial q^2}2\vec{q}\vec{q}+\varphi {\bf U}=\lambda _u{\bf U,}
\end{equation}
and it can easily be seen that 
\begin{equation}
\varphi =\lambda _u
\end{equation}
is a constant, independent of the state variables. Thus we have 
\begin{equation}
s=\lambda _uu+F(q^2),\qquad \vec{J}^S=\lambda _u\vec{q}
\end{equation}

However, a constant temperature is indeed unphysical and furthermore it is
not seen to be related with the temperature $\Theta $ appearing in the
Maxwell-Cattaneo evolution equation, so the previous assumption that $\nabla
\Theta $ is a state function should not be considered. However, if we assume
that $\Theta=\Theta({\cal Z})$, we can write 
\begin{eqnarray}
&&\rho \left( \frac{\partial s}{\partial u}-\lambda _u\right) \frac{\partial
u}{\partial t}+\rho \left( \frac{\partial s}{\partial \vec{q}}-\vec{\lambda}%
_q\right) \cdot \frac{\partial \vec{q}}{\partial t}+\left( \frac{\partial 
\vec{J}^S}{\partial u}-\vec{\lambda}_q\frac{\rho \lambda }\tau \frac{%
\partial \Theta }{\partial u}\right) \cdot \nabla u  \nonumber \\
&&+\left( \frac{\partial \vec{J}^S}{\partial \vec{q}}-\lambda _u{\bf U}-\vec{%
\lambda}_q\frac{\rho \lambda }\tau \frac{\partial \Theta }{\partial \vec{q}}%
\right) :\nabla \vec{q}-\vec{\lambda}_q\frac \rho \tau \vec{q}\geq 0,
\end{eqnarray}
so that, if we define $\xi :=\rho \lambda /\tau $, we obtain 
\begin{equation}
\frac{\partial s}{\partial u}=\lambda _u,\qquad \frac{\partial s}{\partial 
\vec{q}}=\vec{\lambda}_q,
\end{equation}
\begin{equation}
\frac{\partial \vec{J}^S}{\partial u}=\vec{\lambda}_q\xi \frac{\partial
\Theta }{\partial u},\qquad \frac{\partial \vec{J}^S}{\partial \vec{q}}%
=\lambda _u{\bf U}+\vec{\lambda}_q\xi \frac{\partial \Theta }{\partial \vec{q%
}}{\bf ,}
\end{equation}
\begin{equation}
-\vec{\lambda}_q\cdot \frac \rho \tau \vec{q}\geq 0.
\end{equation}

Again we can make use of the representation theorem 
\begin{equation}
\vec{J}^S(u,\vec{q})=\varphi (u,q^2)\vec{q},\qquad \vec{\lambda}_q=\Lambda
(u,q^2)\vec{q},
\end{equation}
\begin{equation}
s=s(u,q^2),\qquad \lambda _u=\lambda _u(u,q^2),\qquad \Theta =\Theta (u,q^2),
\end{equation}
and obtain 
\begin{equation}
\frac{\partial s}{\partial u}=\lambda _u,\qquad \frac{\partial s}{\partial
q^2}=\frac \Lambda 2,  \label{Liu-1}
\end{equation}
\begin{equation}
\frac{\partial \varphi }{\partial u}=\Lambda \xi \frac{\partial \Theta }{%
\partial u},\qquad \frac{\partial \varphi }{\partial q^2}2\vec{q}\vec{q}%
+\varphi {\bf U}=\lambda _u{\bf U}+\Lambda \xi \frac{\partial \Theta }{%
\partial q^2}2\vec{q}\vec{q}.  \label{Liu-2}
\end{equation}
The last equation splits into 
\begin{equation}
\varphi =\lambda _u,\qquad \frac{\partial \varphi }{\partial q^2}=\Lambda
\xi \frac{\partial \Theta }{\partial q^2}.  \label{Liu-3}
\end{equation}
Thus, the entropy flux is seen to adopt the form predicted in (\ref{flux-EIT}%
), that is ($\lambda _u\equiv 1/\theta $), 
\begin{equation}
\vec{J}^S=\frac{\vec{q}}\theta
\end{equation}
From (\ref{Liu-2})$_1$ and (\ref{Liu-3})$_2$, we observe that $\varphi $ is
function of $\Theta $ only, i.e. 
\begin{equation}
\frac{d\varphi }{d\Theta }=\Lambda \xi
\end{equation}
and thus, also $\Lambda \xi $. To proceed, we should make further
assumptions concerning the temperature $\Theta $.

\begin{itemize}
\item  If we assume that $\Theta =\theta $, that is $\varphi =1/\Theta $, we
can write 
\begin{equation}
\Lambda \xi =-\frac 1{\theta ^2}\quad \Rightarrow \quad \Lambda =-\frac
1{\xi \theta ^2}=-\frac \tau {\rho \lambda \theta ^2},
\end{equation}
so we recover EIT's generalized specific entropy: 
\begin{equation}
ds=\frac{du}\theta -\frac \tau {\rho \lambda \theta ^2}\vec{q}\cdot d\vec{q}.
\label{entro}
\end{equation}
The verification of the residual inequality, namely, 
\begin{equation}
\sigma ^S=-\vec{\lambda}_q\frac \rho \tau \cdot \vec{q}\geq 0
\end{equation}
is guaranteed by the negativity of $\Lambda =-\tau /\left( \rho \lambda
\theta ^2\right) $.

Note that no approximation has been performed and therefore, (\ref{entro})
is only limited by the validity of equation (\ref{evo-equ2})$_2$. Thus our
procedure has been quite different from that adopted in usual EIT, where one
departs from a generalized entropy and obtains a Maxwell-Cattaneo-like
equation under the assumptions that a linear relation between fluxes and
forces exists and that $\vec{J}^S=\vec{q}/\theta $. Now, we have seen that
if a Maxwell-Cattaneo evolution equation for the heat flux holds, the
generalized entropy of EIT is the only possible one and the entropy flux
must be given by equation (\ref{flux-EIT}).

\item  If we consider that $\Theta $ is the local-equilibrium temperature $T$%
, it must be independent of the flux $\vec{q}$, so also must be $\theta $,
and, of course, $\theta =T$. Thus we should obtain an specific entropy of
the form 
\begin{equation}
s=s_{eq}(u)+F(q^2)  \label{entropy}
\end{equation}
and $\Lambda =\Lambda \left( q^2\right) $. If no further assumptions are
made, $\xi =\xi (u,q^2)$ and the residual inequality implies, as before that 
$\Lambda \leq 0$. Thus, we observe that we can demand that the ''physical''
temperature appearing in the Maxwell-Cattaneo equation be the
local-equilibrium one and thus, as suggested by Banach in \cite{Banach}, the
correction on entropy due to nonequilibrium situations must be additive and
independent of the equilibrium variables. Let us note, however, that our
requirement is much more restrictive than his. In fact, one of the problems
of developing a extension of CIT\ by spanning the state space is determining
which variables should be considered. One can always choose the
nonequilibrium variables in order that the corrections to entropy do not
depend on the equilibrium quantities. However, if equation (\ref{evo-equ2})
holds and we have already chosen the heat flux $\vec{q}$ as the proper
nonequilibrium variable, we have proven that (\ref{entropy}) must hold. On
the other hand, in \cite{CucaMus} it has also been proposed, within the
context of discrete systems, that the nonequilibrium entropy should take the
form 
\begin{equation}
S=\frac U\Theta +F(\Theta ,\Theta ^{+},\Theta ^{-},\dot{\Theta},\dot{\Theta}%
^{+},\dot{\Theta}^{-}),
\end{equation}
i.e. that out of equilibrium one can consider the generalized contact
temperature as an independent variable (Note that the existence of three
temperatures $\Theta ,\Theta ^{+}$ and $\Theta ^{-}$ is a consequence of
considering a discrete system and replace the temperature field). In this
sense, the correction to the entropy does also depend only on nonequilibrium
variables, although the first term also differs from its equilibrium
counterpart, where $\Theta $ becomes a function of $U$.
\end{itemize}

\section{Heat flux determined by a general evolution equation}

If we now assume an evolution equation for the heat flux as given in
equation (\ref{evo-q}), (note that such evolution equation arises for
instance, when one integrates the Boltzmann equation) and considers that $%
{\bf Q=Q}(u,\vec{q})$ and $\vec{\sigma}^q=\vec{\sigma}^q(u,\vec{q})$, we may
write 
\begin{eqnarray}
&&\rho \left( \frac{\partial s}{\partial u}-\lambda _u\right) \frac{\partial
u}{\partial t}+\rho \left( \frac{\partial s}{\partial \vec{q}}-\vec{\lambda}%
_q\right) \cdot \frac{\partial \vec{q}}{\partial t}+\left( \frac{\partial 
\vec{J}^S}{\partial u}-\vec{\lambda}_q\frac{\partial {\bf Q}}{\partial u}%
\right) \cdot \nabla u  \nonumber \\
&&+\left( \frac{\partial \vec{J}^S}{\partial \vec{q}}-\lambda _u{\bf U}-\vec{%
\lambda}_q\cdot \frac{\partial {\bf Q}}{\partial \vec{q}}\right) :\nabla 
\vec{q}+\vec{\lambda}_q\cdot \vec{\sigma}^q\geq 0,
\end{eqnarray}
so we now have 
\begin{equation}
\frac{\partial s}{\partial u}=\lambda _u,\qquad \frac{\partial s}{\partial 
\vec{q}}=\vec{\lambda}_q,
\end{equation}
\begin{equation}
\frac{\partial \vec{J}^S}{\partial u}=\vec{\lambda}_q\frac{\partial {\bf Q}}{%
\partial u},\qquad \frac{\partial \vec{J}^S}{\partial \vec{q}}=\lambda _u%
{\bf U}+\vec{\lambda}_q\cdot \frac{\partial {\bf Q}}{\partial \vec{q}}{\bf ,}
\end{equation}
\begin{equation}
-\vec{\lambda}_q\cdot \vec{\sigma}^q\geq 0.
\end{equation}
Once more can we make use of a representation theorem to write 
\begin{eqnarray}
\vec{J}^S(u,\vec{q}) &=&\varphi (u,q^2)\vec{q},\qquad \vec{\lambda}%
_q=\Lambda (u,q^2)\vec{q},\qquad \vec{\sigma}^q=\sigma (u,q^2)\vec{q}, \\
s &=&s(u,q^2),\qquad \lambda _u=\lambda _u(u,q^2),\qquad {\bf Q}=a(u,q^2)%
{\bf U}+b(u,q^2)\vec{q}\vec{q},
\end{eqnarray}
so the previous equations may be simplified and one obtains 
\begin{eqnarray}
\frac{\partial s}{\partial u} &=&\lambda _u,\qquad \frac{\partial s}{%
\partial q^2}=\frac \Lambda 2,  \label{Liu-21} \\
\frac{\partial \varphi }{\partial u} &=&\Lambda \left( \frac{\partial a}{%
\partial u}+\frac{\partial b}{\partial u}q^2\right) ,\qquad \frac{\partial
\varphi }{\partial q^2}=\Lambda \left[ \frac{\partial a}{\partial q^2}+\frac{%
\partial b}{\partial q^2}q^2+\frac b2\right] ,  \label{Liu-22} \\
\varphi &=&\lambda _u+\Lambda bq^2.  \label{Liu-23}
\end{eqnarray}
Let us observe that the entropy flux we obtain is given by (we write $%
\lambda _u=1/\theta $): 
\begin{equation}
\vec{J}^S=\frac{\vec{q}}\theta +\Lambda bq^2\vec{q},
\end{equation}
while in \cite{Cuca} we had obtained 
\begin{equation}
\vec{J}^S=\frac{\vec{q}}\theta +\Lambda \left( a+bq^2\right) \vec{q}.
\end{equation}
Thus we can observe that in this concrete example equation (\ref{malfet})
does not hold. Note, in addition that, in equilibrium, ${\bf Q}$ is not null
but isotropic, thus this latter expression will also differ from the usual
local-equilibrium one, namely, $\vec{J}^S=\vec{q}/T$ and thus is not
convenient. If we define 
\begin{equation}
m:=a+bq^2,\qquad x:=\frac{q^2}2,\qquad \Lambda b:=f,
\end{equation}
equations (\ref{Liu-21}) and (\ref{Liu-22}) may be simplified and yield

\begin{equation}
\frac{\partial s}{\partial u}=\varphi -2xf,\qquad \frac{\partial s}{\partial
x}=\Lambda ,  \label{68}
\end{equation}
\begin{equation}
\frac{\partial \varphi }{\partial u}=\Lambda \frac{\partial m}{\partial u}%
,\qquad \frac{\partial \varphi }{\partial x}=\Lambda \frac{\partial m}{%
\partial x}-f.  \label{69}
\end{equation}

In order to obtain a concrete expression for the entropy of the system, we
must solve this set of equations, taking into account that we must recover
the equilibrium result in the case of null heat flux together with the
convexity requirement for the entropy. Such requirement, namely, that $%
\delta ^2s\geq 0$ yields the following inequalities: 
\begin{eqnarray}
&&\frac{\partial \varphi }{\partial u}-2x\frac{\partial f}{\partial u}\leq
0,\qquad \Lambda \leq 0,\qquad \Lambda +\frac{\partial \Lambda }{\partial x}%
2x\leq 0, \\
&&\left( \frac{\partial \varphi }{\partial u}-2x\frac{\partial f}{\partial u}%
\right) \left( \Lambda +\frac{\partial \Lambda }{\partial x}2x\right)
-\left( \frac{\partial \Lambda }{\partial u}\right) ^22x\geq 0.  \nonumber
\end{eqnarray}
If we restrict ourselves up to second order in the heat flux $\vec{q},$ as
both $\Lambda $ and $b$ vanish in local-equilibrium, we may take $f=0,$ so
equations (\ref{68}) and (\ref{69}) reduce to 
\begin{equation}
\frac{\partial s}{\partial u}=\varphi ,\qquad \frac{\partial s}{\partial x}%
=\Lambda ,  \label{68b}
\end{equation}
\begin{equation}
\frac{\partial \varphi }{\partial u}=\Lambda \frac{\partial m}{\partial u}%
,\qquad \frac{\partial \varphi }{\partial x}=\Lambda \frac{\partial m}{%
\partial x},  \label{69b}
\end{equation}
and within this approximation the entropy flux reduces to (\ref{flux-EIT})
and $m$ is a function of $\varphi $.

\section{Concluding remarks}

In this paper we have applied Liu's technique to the simple situation
considered in \cite{Cuca}, where EIT\ and Information Theory are applied to
a system submitted to a heat flux $\vec{q}.$ Such a simplification allows to
go beyond a linear approximation. We observe how by using Liu's technique,
one can recover the results in usual EIT, but the procedure is quite
different. As shown in \cite{Cuca}, in EIT one assumes a flux-dependent
specific entropy and a concrete form for the entropy flux, namely, $\vec{J}%
^S=\vec{q}/\theta $, and thus one is led to a hyperbolic evolution equation
for the heat flux. Now, the viewpoint is the opposite: we depart from a
given evolution equation for the heat flux, and are thus led to a concrete
form for the specific entropy an the entropy flux.

On the other hand, we have considered a general evolution equation for the
heat flux, namely 
\begin{equation}
\frac{\partial \vec{q}}{\partial t}+\nabla \cdot {\bf Q}=\vec{\sigma}^q,
\end{equation}
and derived the corresponding form for the entropy flux and the set of
partial differential equations whose solution allows the determination of
the generalized nonequilibrium entropy of the system. We have also observed
that our results differ from those obtained in \cite{Cuca}, due to the
definition employed for the entropy and entropy flux. Our present result,
namely, 
\begin{equation}
\vec{J}^S=\vec{q}/\theta +\Lambda bq^2\vec{q},
\end{equation}
which reduces to $\vec{J}^S=\vec{q}/T$ in local-equilibrium, is thus much
more convenient.

Hence, we observe that there is not a complete equivalence between Liu's
technique and information theory. This lack of equivalence is also observed
when discrete systems are considered \cite{CucaMus}. For such systems, one
cannot usually write evolution equations for the fluxes, so by applying
Liu's technique it is possible to prove that entropy cannot depend on them.
On the other hand, if one assumes that the state of the system is described
by the total internal energy $U$ and the heat flux $\dot{Q},$ information
theory yields to an entropy which depends on both $U$ and $\dot{Q},$ in
spite of the fact that no evolution equations neither for $U$ nor for $\dot{Q%
}$ need to be included. This is due to the fact that we deal with such
restrictions by maximizing 
\begin{equation}
-k_B\int f\ln fd\vec{c}-\lambda _U\int fHd\vec{c}-\lambda _Q\int f\hat{Q}d%
\vec{c}.
\end{equation}
Thus, we see that there may be some subtle differences between the use of
Lagrange multipliers in usual information theory and in Liu's procedure.

\section*{Acknowledgments}

Stimulating discussions with Prof. J. Casas-V\'{a}zquez and J. Faraudo are
acknowledged. One of the authors (R.D-C) is supported by a doctoral
scholarship from the Programa de formaci\'{o} d'investigadors of the
Generalitat de Catalunya under grant FI/94-2.009. We also acknowledge
partial financial support from the Direcci\'{o}n General de
Investigaci\'{o}n of the Spanish Ministry of Education and Science (grant
PB94-0718).

\end{document}